\documentclass[reprint,aps,prb,twocolumn,floats,footinbib,superscriptaddress, longbibliography]{revtex4-1}
\usepackage{times}
\usepackage[outdir=./]{epstopdf}
\usepackage[T1]{fontenc}
\usepackage{bm}
\usepackage{graphicx}
\usepackage{amssymb}
\usepackage{leftidx}
\usepackage{upgreek}
\usepackage{siunitx}
\usepackage{amsfonts}
\usepackage{amsmath} 
\usepackage{color}
\usepackage{float}
\usepackage{verbatim}
\usepackage{fancyvrb} 
\usepackage{wasysym}
\usepackage{hyperref}

\usepackage{xspace}
\newcommand{\pp}{\ensuremath{^{2+}}\xspace}

\usepackage{balance}

\begin{document}
\title{\large \textbf{Strain-Induced Speed-Up of Mn\pp Spin-Lattice Relaxation in (Cd,Mn)Te/(Cd,Mg)Te Quantum Wells: A Time-Resolved ODMR Study}}

\def \FUW{Institute of Experimental Physics, Faculty of Physics, University
of Warsaw, ul. Pasteura 5, 02-093 Warsaw, Poland}

\author{A. \surname{Bogucki}}\affiliation{\FUW}
\author{A. \surname{{\L}opion}}\affiliation{\FUW}
\author{K. E. \surname{Po{\l}czy{\'n}ska}}\affiliation{\FUW}
\author{W. \surname{Pacuski}}\affiliation{\FUW}
\author{T. \surname{Kazimierczuk}}\affiliation{\FUW}
\author{A. \surname{Golnik}}\affiliation{\FUW}
\author{P. \surname{Kossacki}}\affiliation{\FUW}

\date{\today}

\begin{abstract}
This study examines the spin-lattice relaxation rate of Mn\pp ions in strained diluted magnetic semiconductor (Cd,Mn)Te/(Cd,Mg)Te quantum wells using the optically detected magnetic resonance (ODMR) technique. By adjusting the magnesium (Mg) content in the buffer layer, we created samples with different strain levels. Our time-resolved ODMR results show that the spin-lattice relaxation time becomes faster as strain increases. We also found that the relaxation rate increases with both magnetic field and temperature, showing a power-law behavior.
To understand these observations, we used a theoretical model based on six-level rate equations with non-equal level separations. This model suggests that the main factor affecting relaxation in our samples is a "direct" mechanism. The model's predictions match well with our experimental data.
Overall, our findings give insights into spin-lattice relaxation in strained quantum wells and could be important for the development of future quantum and spintronic devices.
\end{abstract}


\maketitle

\section{INTRODUCTION}
Understanding spin-lattice relaxation mechanisms is of paramount importance for potential applications in quantum and spintronic devices\cite{Norambuena_2018_PRB, Debus_2016_PRB, Koenraad_2011_NM, Preskill_1998_PRSLA}. In this study, we aim to investigate the impact of strain in the crystal lattice surrounding magnetic ions on their spin relaxation. Our focus is on (Cd,Mn)Te/(Cd,Mg)Te quantum wells with  incorporated low density of Mn\pp ions, which provide an intermediate system between quantum dots and bulk crystals, enabling us to explore the strain dependence.

\begin{table}[h]
\caption{Composition of samples used in this article and deformation
$\varepsilon_{\parallel}$
present in QW layer.}
\begin{tabular}{l|c|c|c|r}
\toprule
sample no. &  buffer Mg (\%) & barrier Mg (\%) & QW Mn (\%) &
$\varepsilon_{\parallel}$(\permil)    \\ \hline
UW1029     & 30.7           & 30.7            & 0.30        & $-$3.21     \\
UW1030     & 21.2           & 21.2            & 0.31       & $-$2.31     \\
UW1031     &  0.0           & 21.2            & 0.26       & $-$0.32     \\
\hline \hline
\end{tabular}
\label{tab:samples}
\end{table}

\begin{figure}[!th]
\begin{center}
\includegraphics[width=0.9\columnwidth]{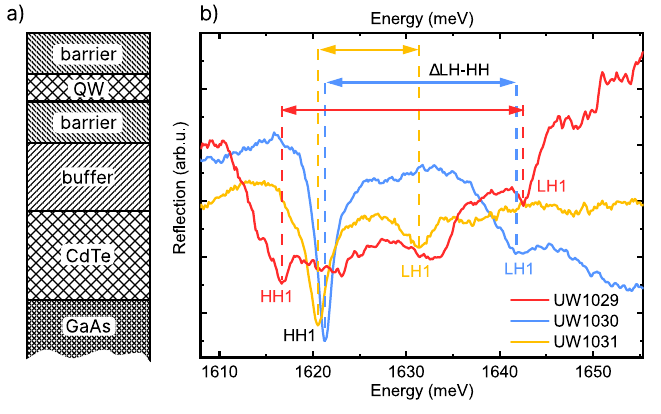}
\end{center}
\caption[]{(Color online) a) A representative sample structure, used across
the series of samples in this article. The buffer layer is composed of (Cd,
Mg)Te, with magnesium content that varies from 0\% to 30.7\%. Barriers were
constructed from (Cd, Mg)Te, while the quantum well (QW) was made from (Cd,
Mn)Te, and contained a manganese concentration of less than 0.5\%. b) The
reflectivity spectra of studied samples demonstrate the characteristic
features corresponding to the heavy-hole exciton (HH) and light-hole
exciton (LH). The energy splitting between these two ($\Delta$LH-HH) is
related to the strain present in the QW layer. The larger the
$\Delta$LH-HH, the larger the strain. }
\label{fig:F1_ODMR_relax_Refl_Spectra}
\end{figure}

When studying crystals doped with magnetic ions, it is crucial to consider that even at very low dopant concentrations, a single paramagnetic ion or a spin system cannot be treated as completely isolated from its environment\cite{Larson_1988_PRB}. The investigated system experiences constant energy exchange processes between the magnetic ions and the generalized thermal reservoir\cite{Standley_1969_book}. This reservoir comprises lattice phonon vibrations (phonon excitations of the crystal), carriers, and the crystal environment (e.g., helium bath), all of which can affect the availability of relaxation pathways. Consequently, an ion in an excited state (e.g., due to the absorption of radiation) gradually loses the previously absorbed energy over time and returns to the equilibrium. The most common relaxation process in our system is the direct process mechanism \cite{Abragam_2012_EPR}, also known as the spin-flip process. This process involves a change in the spin projection (transition between energy levels resulting from an external magnetic field splitting due to the emission or absorption of a single phonon from or into the crystal lattice \cite{Culvahouse_1963_PR}.

Spin-lattice relaxation has been extensively studied in various diluted magnetic semiconductor (DMS) systems. These include bulk crystals \cite{Strutz_1992_PRL, Strutz_1993_PRB, Witowski_1995_PBCM}, quantum well systems \cite{Yakovlev_2004_PSSC, Ivanov_2008_PRB}, and lower-dimensional nanostructures \cite{Hundt_2005_PRB, Tolmachev_2020_N}. A range of techniques has been employed to investigate the time-dependent magnetic properties of these systems. These include infrared laser pulse induced electron spin resonance \cite{Strutz_1993_PRB}, photocarrier-induced heating effects in quantum dots \cite{Hundt_2005_PRB}, phonon-induced variations in photoluminescence intensity \cite{Scherbakov_2004_pss}, and time-resolved Faraday rotation \cite{Crooker_1997_PRB}. Among these methods, Optically Detected Magnetic Resonance (ODMR) \cite{Geschwind_1959_PRL, Komarov_1977_JETP, Godlewski_2008_OM, Ivanov_2008_PRB, Tolmachev_2012_JL} stands out. Its high spatial resolution and sensitivity are particularly advantageous for our study, as they allow us to specifically target and analyze the small volume of the sample where strain and magnetic ions are present.
\newpage

\section{SAMPLES, EXPERIMENTAL SETUP AND METHODS}
\subsection{Samples with controlled deformation in QW layer}
The samples studied in this paper consist of quantum wells (QWs) created by
the molecular beam epitaxy (MBE) method. Different strains in the samples
(as well as deformation $\varepsilon_{\parallel}$)  were achieved by
modifying the magnesium content in the buffer and barrier layers. Detailed
characterization of the presented samples (UW1029, UW1030, UW1031) can be
found in the article \cite{Bogucki_2022_PRB}, therefore here we describe
them only briefly. The representative scheme for all used samples is
illustrated in Fig. \ref{fig:F1_ODMR_relax_Refl_Spectra}a), and the
composition of each layer is described in Table \ref{tab:samples}. The
samples were grown on GaAs substrates, and a CdTe layer of 4~$\upmu$m
thickness was grown on top of the substrate for decoupling from the GaAs
lattice constant\cite{Cibert_1990_APL, Cibert_1991_SaM}. A (Cd, Mg)Te buffer layer of 2~$\upmu$m (larger than the
critical thickness of (Cd, Mg)Te \cite{Waag_1993_JoCG, Gerthsen_1994_JoAP}) was used to govern the strain in the
samples, and thin (Cd, Mg)Te barriers of 50~nm (smaller than the critical
thickness) were incorporated. We have determined the magnesium content
using magnesium fluxes and reflectance measurements
\cite{Hartmann_1996_JoAP, LeBlanc_2017_JoEM}. The thickness of the (Cd,
Mn)Te quantum wells (QWs) was selected at 10~nm, which is well below the
critical thickness of CdTe lattice relaxation \cite{Cibert_1990_APL,
Cibert_1991_SaM}. Such a choice ensured good confinement and the appearance
of narrow excitonic features. The modified Brillouin function was fitted to
the giant Zeeman splitting to verify the manganese (Mn) content
\cite{Gaj_1994_PRB}. We identified the reflectance spectrum features,
including the assignment of heavy-hole excitonic states and light-hole
excitonic states, using standard magneto-optical measurements
\cite{Kossacki_1997_SSC, Kossacki_2003_JPCM, Kossacki_2004_PRB}. The
reflectivity spectra of the samples, revealing identified features, are
shown in Fig. \ref{fig:F1_ODMR_relax_Refl_Spectra}b. The energy of the spectrum features confirms the
designed strain in the samples. Moreover, XRD measurements have further
verified the corresponding deformation values as described in~\onlinecite{Bogucki_2022_PRB}.

\subsection{Time-resolved ODMR}

Optically Detected Magnetic Resonance (ODMR) is a method based on the interaction of photo-created carriers and magnetic ions in a sample. Microwave radiation of constant frequency is delivered to the sample, placed within a constant external magnetic field. When the energy at a specific microwave frequency resonates with a paramagnetic transition at a particular magnetic field, paramagnetic resonance occurs, leading to a decrease in the net magnetization of the sample.

The decrease in magnetization, caused by paramagnetic resonance, is detectable as a change in the optical spectrum due to the exchange interaction between photo-generated carriers and magnetic spins. The ODMR method offers spatial resolution and sensitivity, probing only a small part of the sample (excitation laser was focused on the sample resulting in spot with approximately 100~$\mu$m in diameter) and the thin Quantum Well (QW) where the magnetic ions are present (10~nm).

The difference in the optical spectrum at resonant conditions, compared to the spectrum measured without microwave radiation, is defined as the ODMR signal. In this study, time-resolved ODMR was used, employing pulsed microwave radiation and pulsed optical excitation/probing of samples. The microwave pulse period was 14~ms with a pulse width of 2~ms, and the optical probing pulse had the same period of 14~ms with a duration of 0.5~ms. 

\onecolumngrid

\begin{center}
\rule{0.3\textwidth}{0.4pt}
\end{center}
\vspace*{-0.5cm}
\begin{figure}[!hb]
\begin{center}
\includegraphics[width=1\columnwidth]{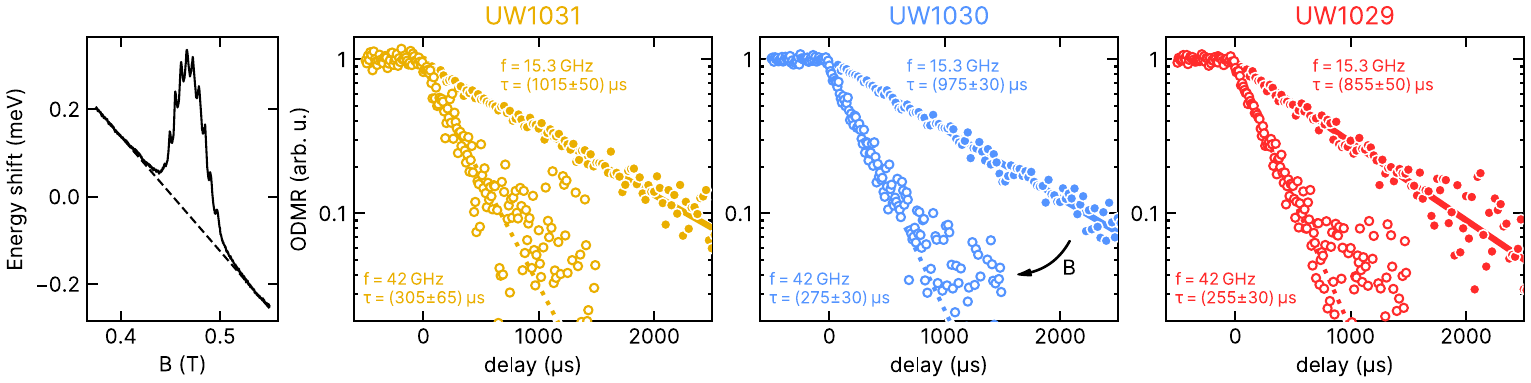}
\end{center}
\caption[]{(Color online) The left panel presents the ODMR spectrum (solid line) of CdTe quantum well doped with manganese ions (sample UW1031), measured in a presence of microwave radiation ($f=13.175$~GHz). The y-axis represents the energy of the recombination of the neutral exciton complex (X), as visible in the photoluminescence spectrum, with an offset of $1619.7$~meV subtracted. The dashed line corresponds to the position of X without microwave radiation. The ODMR signal is defined as the difference between both lines at resonant magnetic field. Other panels show time evolution of the ODMR signal as a function of the time separating the microwave radiation pulse and the probing light pulse, measured in CdTe quantum wells doped with manganese ions. Solid points and lines correspond to the microwave frequency $f=15.3$~GHz. Hollow points and dashed lines correspond to the microwave frequency $f=42$~GHz. The lines represent exponential decay fit curves with the characteristic time $\uptau$. The three panels show the results obtained for three quantum wells with different nominal strain values, with the strain increasing from the left panel to the right. The determined values of the parameter $D$ and the strain are presented in Figure \ref{fig:F4_ODMR_relax_vs_Deformation} }
\label{fig:F2_ODMR_relax_curves}
\end{figure}

\clearpage
\newpage

\twocolumngrid
\noindent By changing the delay between the microwave and light pulses, the relaxation curve of the magnetic ions could be reconstructed. This method allows for the measurement of the dynamics of the ODMR signal, and therefore the spin-lattice relaxation rate of Mn\pp ions in CdTe QWs. During the experiments excitation power density of the laser was below 0.01~W/cm$^2$ -- low enough to avoid heating the lattice of the sample, as evidenced by lack of detectable redshift of the spectrum\cite{Keller_2001_PRB}.

A well-established method of obtaining high values of microwave radiation at the position of the sample is by using a microwave cavity, commonly employed in EPR measurements. However, this approach is not ideal for optical studies, as optical access to the sample becomes challenging. Moreover, the resonant microwave cavity is typically optimized for a limited number of frequencies. Therefore, in this work, we utilize a sample holder based on printed circuit boards, which effectively creates a closed loop of copper ribbon around the sample. This ensures the correct orientation (perpendicular) of the magnetic component of the microwave radiation with respect to the static magnetic field and the optical axis direction. The employed sample holder yields a robust ODMR signal across a wide range of microwave frequencies -- we have tested it from 12 to 42~GHz.

\section{RESULTS}
\vspace*{-1ex}
\subsection{Time-resolved ODMR measurements}
\vspace*{-3.5mm}
FIG. \ref{fig:F2_ODMR_relax_curves} presents the ODMR relaxation curves obtained for the measured samples, where deformation present in the QW increases from the left panel (sample UW1031) to the right panel (sample UW1029). The horizontal axis depicts the delay between the MW pulse and the light probing pulse, whereas the vertical axis represents the normalized ODMR signal amplitude. Each panel consists of two decay curves: the filled circles correspond to a MW frequency of 15.3~GHz, which is in tune with the resonant magnetic field of 0.547~T. The empty circles were obtained at a MW frequency of 42~GHz, resonating with a magnetic field of 1.5~T. It is evident that the relaxation time is shorter at higher magnetic fields (characteristic decay time approximately 0.3~ms at a higher magnetic field and around 1~ms for a lower magnetic field) for all samples. The results showcased in FIG. \ref{fig:F2_ODMR_relax_curves} were recorded at a temperature of 1.6~K (pumped helium bath). 

\onecolumngrid

\begin{center}
\rule{0.3\textwidth}{0.4pt}
\end{center}
\vspace*{-0.5cm}
\begin{figure}[!htb]
\begin{center}
\includegraphics[width=0.99\columnwidth]{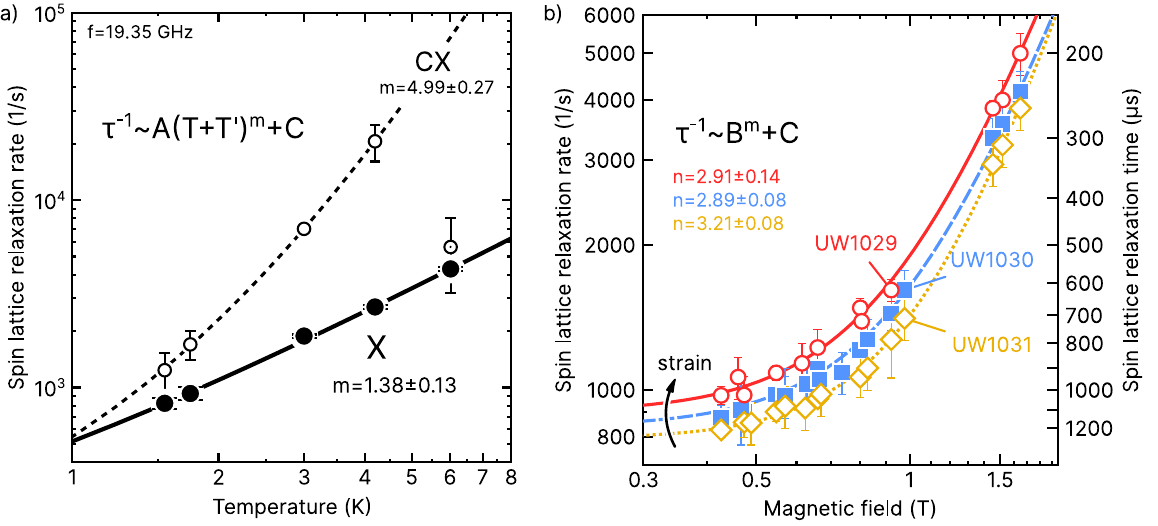}
\end{center}
\caption[]{(Color online) Spin-lattice relaxation rates of (Cd, Mn)Te/(Cd, Mg)Te QWs determined from time-resolved ODMR measurements. Left panel (a) shows spin-lattice relaxation rate vs temperature extracted from ODMR signal detected on neutral exciton line (X, marked with black dots) and charged exciton line (CX, marked with empty circles). The solid and dashed lines represent best fit of $\uptau^{-1}=A(T+T')^m$ formula to the data. The determined exponents are $m=(1.38 \pm 0.13)$ for neutral exciton and $m=(4.99\pm0.27)$ for charged exciton. The empty circle marked with a black dot inside was excluded from fitting due to the low signal-to-noise ratio.  Magnetic field dependence of spin-lattice relaxation rates is shown on panel (b). Symbols represent experimental data, and the solid lines represent the fitted power function of the form: $\text{SLRR} \sim B^n + C$ with exponents: $n = (2.91 \pm 0.14)$ for sample UW1029, $n = (2.89 \pm 0.08)$ for sample UW1030, and $n = (3.21 \pm 0.08)$ for sample UW1031. Both the measured points and the fitted lines follow the trend that the greater the deformation in the quantum well, the faster the ion relaxation rate, meaning a shorter relaxation time $\uptau$.}
\label{fig:F3_ODMR_relax_Gorycogram_and_T}
\end{figure}
\newpage

\twocolumngrid

Since the relaxation rate is highly temperature-dependent, we measured its variation with temperature for sample UW1467, which contains 23\% of Mg in the barrier and buffer, and 0.27\% of Mn in the QW. The UW1467 sample bears a close resemblance in construction to UW1029. It was selected for temperature-dependent measurements owing to its exceedingly narrow excitonic line of approximately 1.2~meV. Minimizing line width is crucial because of line broadening at elevated temperatures; starting with the narrowest possible lines facilitated the identification of the ODMR signal at higher temperatures. The acquired spin-relaxation rates are depicted in the left panel of FIG. \ref{fig:F3_ODMR_relax_Gorycogram_and_T}. The observed spin-lattice relaxation rate temperature dependence aligns well with the power law $\uptau^{-1}=A(T+T')^{m}+C$, where $T'=0~K$ and the exponent $m=(1.38\pm0.1)$ is ascertained for the neutral exciton, implying that the relaxation is governed by a "direct" mechanism involving a single phonon since the exponent is near 1. A similar analysis can be performed for the ODMR signal derived from the charged exciton line. In this case to achieve a satisfactory fit, the phenomenological temperature shift of $T'=(2\pm0.3)~K$ was added, eventually deriving an exponent for the charged exciton as $m=(5\pm0.3)$. Due to the low signal-to-noise ratio, the 6~K point sourced from the charged exciton line (open circle marked with a black dot inside) was excluded from the fitting process. The difference in the measured rate has profound consequences, proving that magnetic ions probed by the charged exciton are not the same as those probed by the neutral exciton and that carriers plays an important role in the relaxation process. This observation will be elaborated upon in the \ref{sec:SUMMARY} section.

Additional information about the spin-lattice interaction may be retrieved from time-resolved ODMR measurements at different values of external magnetic field. The range of magnetic fields that were used during the experiments was limited by our microwave setup that covers frequency ranges of 12--27~GHz and 42--47~GHz which corresponds to resonant magnetic fields (for typical Mn\pp g-factor) 0.428--0.963~T and 1.5--1.68~T respectively.

The spin-lattice relaxation rates as a function of resonant magnetic field for measured samples are presented on right panel of FIG. \ref{fig:F3_ODMR_relax_Gorycogram_and_T} with empty symbols. Obtained spin-lattice relaxation rate dependencies are well described by formula $\uptau^{-1}=\alpha B^{n}+\uptau_{0}^{-1}$, where $\alpha$, $n$ and $\uptau_0^{-1}$ are free parameters. The extracted exponents yield $n=(2.91\pm 0.14)$, $n=(2.89\pm 0.08)$, $n=(3.21\pm 0.08)$ for samples UW1029, UW1030 and UW1031, respectively. All exponents are close to the value 3. Similar dependencies discussed in the literature provide $n$ values of 2, 4, 5 for different ions and experimental conditions \cite{Baker_1964_PR, Strutz_1992_PRL}. At this stage, obtained in this work experiment exponent 3 is not easy to interpret. However, as it is shown in section \ref{subsec:model}, this value in combination with numerical modelling of the system, gives additional evidence that the main spin-lattice relaxation mechanism involves only one phonon. 

\begin{figure}[!th]
\begin{center}
\includegraphics[width=0.99\columnwidth]{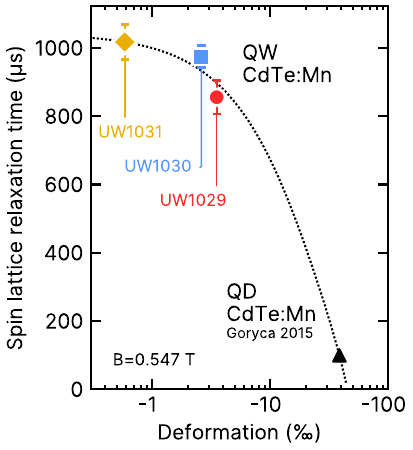}
\end{center}
\caption[]{(Color online) The relaxation time $\uptau$ of the Mn\pp ion in CdTe determined from ODMR measurements of QWs is shown as a function of the deformation present in the QW. The result was obtained for a microwave radiation frequency of $f=15.3$~GHz ($B=0.547$~T). As a reference, the value determined for a CdTe quantum dot with a single manganese ion (data from publication \onlinecite{Goryca_2015_PRB} interpolated to the magnetic field value $B=0.547$~T) is added and indicated on the plot with the symbol $\blacktriangle$. The dotted line is added as a guide for the eye. The relaxation time decreases as the strength of the crystal lattice deformation around the Mn\pp ion increases.}
\label{fig:F4_ODMR_relax_vs_Deformation}
\end{figure}

Interestingly, the fitted curves (red line, blue dashed line, orange dotted line) reveal systematic shift following the strain value present in quantum well layer.  This is more clearly seen in  FIG. \ref{fig:F4_ODMR_relax_vs_Deformation} where spin-lattice relaxation time is presented as a function of deformation that is present in the QW layer. The thin dotted line is added as a guide for the eye. The spin-lattice relaxation is faster for Mn\pp ions incorporated in more strained CdTe material. The black triangle is spin-lattice relaxation time obtained for the CdTe/ZnTe quantum dot containing single Mn\pp~ion\cite{Goryca_2015_PRB}. In such a quantum dot the deformation is calculated as lattice mismatch between the CdTe and ZnTe materials. 

\subsection{Numerical simulations} \label{subsec:model}
\begin{figure}[!th]
\begin{center}
\includegraphics[width=0.99\columnwidth]{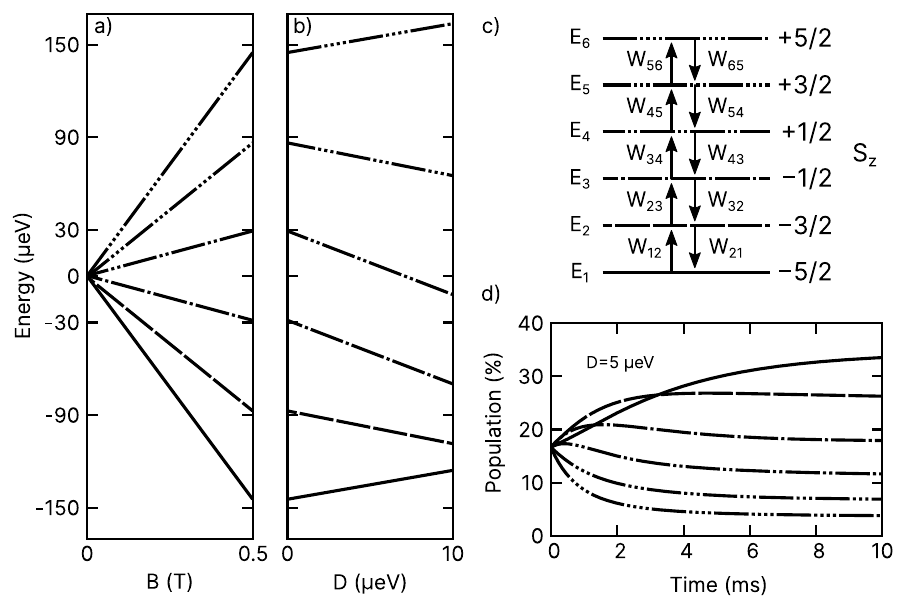}
\end{center}
\caption[]{a) Zeeman splitting of 6-level Mn\pp system (neglecting nuclear spin) vs magnetic field. b) The same 6-level system at constant magnetic field of 0.5 ~T interacting with strained crystal lattice -- horizontal axis presents added ZFS term described by the strain related spin Hamiltonian parameter $D$. The distance between energy levels can be derived from eq. \ref{eq:energy_of_simple_six_levels}. c) Assignment of the electron spin projections on the quantization axis to the different energy levels and designation of transitions between levels. d) Temporal evolution of energy levels populations shows that the lowest energy level population (solid line) increases with time. The mean spin at each time moment is proportional to the magnetization present in  QW.}
\label{fig:F5_ODMR_relax_levels_occup_simulations_Wit}
\end{figure}

\vspace*{-0.5cm}
We aim to understand the observed data through modeling the spin-lattice relaxation in (Cd,Mn)Te/(Cd,Mg)Te quantum wells, using an adapted version of A. Witowski's model \cite{Witowski_1992_APPA}. The main assumption of this model is the "direct" relaxation mechanism, where only one phonon is involved, neglecting weak interactions with the nuclear spin. This simplifies the Mn\pp~system to six energy levels corresponding to the electron spin projections on the quantization axis determined by the magnetic field, with deformation effects reflected by the axial term of the spin-Hamiltonian (parameter $D$ -- see FIG. \ref{fig:F5_ODMR_relax_levels_occup_simulations_Wit}a and FIG. \ref{fig:F5_ODMR_relax_levels_occup_simulations_Wit}b). The simplified spin Hamiltonian is given by:

\begin{equation}
\widehat{H}=g_{\mbox{\tiny{Mn}}}\mu_B\mathbf{\hat{B}\hat{S}}+D\left[\hat{S}_z^2-\frac{S(S+1)}{3}\right],
\label{eq:hamiltonian}
\end{equation} 
where manganese electronic spin $S=5/2$, $\mu_B$ is Bohr magneton,
$g_{\mbox{\tiny{Mn}}}$ is g-factor of manganese ion, and $D$ is the axial zero-field splitting (ZFS) parameter ($D$ is also called the strain-induced axial-symmetry parameter)\cite{Abragam_2012_EPR}. Parameter $D$ can be translated into deformation using spin-lattice coupling parameter ${G_{11}=(72.2\pm1.9)}$~neV, as presented in ref. \onlinecite{Bogucki_2022_PRB}. In the case of studied QWs the sign of deformation is opposite to the sign of the spin Hamiltonian $D$ parameter \cite{Qazzaz_1995_SSC}.

The kinetic model describes the temporal evolution of the occupancies of the six energy levels. The rate coefficients $W_{ij}$ and $W_{ji}$ governing the transitions between levels are derived from the energy differences between the levels and Boltzmann factors  -- see FIG. \ref{fig:F5_ODMR_relax_levels_occup_simulations_Wit} c) -- with expressions for $W_{ij}$ and $W_{ji}$ given as:

\begin{align}
W^{\uparrow}_{ij}&=A_{|j-i|} (|E_{j}-E_{i}|)^3 \frac{1}{e^{\frac{|E_{j}-E_{i}|}{k_b T}}-1}\\
W^{\downarrow}_{ji}&=A_{|j-i|} (|E_{j}-E_{i}|)^3 \left( 1+ \frac{1}{e^{\frac{|E_{j}-E_{i}|}{k_b T}}-1} \right).
\label{eq:kinetic}
\end{align}

Here, $E_{i}$ represents the energy of level $i$, and $k_b T$ is the Boltzmann factor associated with temperature $T$, $A_{|j-i|}$ are numerical constants \cite{Witowski_1993_SSC}. The energy spacing between levels $i+1$ and $i$ is given by:

\begin{equation}
\Delta E_{i+1,i} =g_{\mathrm{Mn}}\mu_B B + 2(i-3)D.
\label{eq:energy_of_simple_six_levels}
\end{equation}

We consider only dipole transitions, which change spin projection by $\Delta S_z=1$. An example of the evolution of level populations is presented in FIG. \ref{fig:F5_ODMR_relax_levels_occup_simulations_Wit} d). We then calculate the evolution of the mean spin value's, which is approximated by a monoexponential decay \cite{Witowski_1992_APPA}, with the characteristic decay time being the spin-lattice relaxation time.

\begin{figure}[!th]
\begin{center}
\includegraphics[width=0.9\columnwidth]{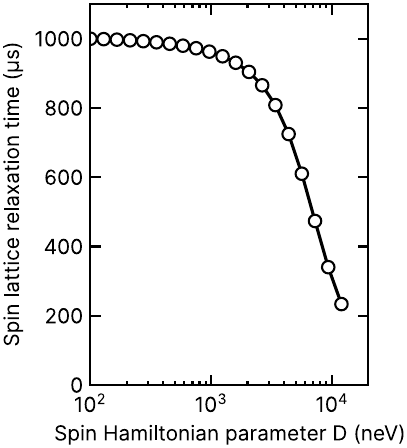}
\end{center}
\caption[]{ Spin-lattice relaxation of Mn\pp ion in CdTe strained QW vs strain-related spin Hamiltonian parameter $D$. Each dot represents an exponential-decay-fit to the temporal evolution of mean spin calculated from curves similar to the presented in FIG. \ref{fig:F5_ODMR_relax_levels_occup_simulations_Wit}d). Obtained result is in qualitative agreement with data presented in FIG. \ref{fig:F4_ODMR_relax_vs_Deformation}}
\label{fig:F6_ODMR_relax_vs_D_simulation}
\end{figure}

Values of spin-lattice relaxation times obtained, from such procedure, for multiple values of strain-related spin Hamiltonian parameter $D$ are shown in FIG. \ref{fig:F6_ODMR_relax_vs_D_simulation}. The spin-lattice relaxation time becomes shorter with increasing strain-related spin Hamiltonian parameter $D$. This behavior is in good agreement with measured dependence presented in FIG. \ref{fig:F4_ODMR_relax_vs_Deformation}. Moreover, similar simulations were performed as a function of the lattice temperature and magnetic field. Obtained dependencies of spin-relaxation rate are also in agreement with measured data and are presented in appendix (FIG. \ref{fig:S1_ODMR_relax_vs_T_simulations} and FIG. \ref{fig:S2_ODMR_relax_vs_B_simulations}). 

\section{DISCUSSION, SUMMARY and CONCLUSIONS} \label{sec:SUMMARY}

It's worth noting that the sensitivity of ODMR measurements is inherently temperature-dependent, as the ODMR signal is based on the Giant Zeeman effect. The amplitude of the Giant Zeeman effect decreases as the temperature of the system rises -- this is a result of the decreasing net magnetization of the sample. Therefore, the practical upper limit of temperatures at which ODMR was measurable in our experimental setup was close to 6~K. The maximum ODMR amplitude for the neutral excitonic line at 6~K was 50~$\upmu$eV. The analogous value for the signal extracted from the charged exciton was 30~$\upmu$eV with a standard deviation of 8.5~$\upmu$eV. The obtained temperature dependence of the spin-relaxation rate for the neutral exciton strongly suggests a one-phonon relaxation mechanism. The spin-relaxation rate obtained from the charged exciton shows a much stronger dependence on temperature. One explanation for this observation could be that charged excitons are probing different populations of manganese ions. One can imagine local, submicrometer-scale spatial fluctuations of potential, where a charged exciton forms more frequently. The manganese ions located in such a charged area may have more possibilities of transferring energy to the crystal-lattice, as more quasiparticles are involved. Therefore, in case of manganese ions interacting with charged exciton the temperature exponent $m$ of the spin-lattice relaxation rate may be greater than 1. During the experiments laser spot had a diameter of approximately 100~$\upmu$m resulting in clear visibility of both charged and neutral excitons. In our case both the charged and neutral exciton exhibited resonance at the same magnetic field -- we do not observe excitonic Knight shift \cite{Story_1996_PRL, Konig_2000_PRB, Teran_2003_PRL, Konig_2003_PRL}. 

The model presented in section \ref{subsec:model}, which assumes a one-phonon scattering mechanism, reproduces all observed trends: spin-lattice relaxation vs temperature, vs magnetic field, and vs deformation. However, the obtained exponent $n\approx3$ is not common in the literature in the context of the direct spin-lattice relaxation mechanism. This discrepancy can be attributed to the energy regimes at which the measurements were taken. In this work, the energy splittings corresponding to the paramagnetic resonance and zero field splitting are of the same order as the thermal energy. This is evident in FIG. \ref{fig:F5_ODMR_relax_levels_occup_simulations_Wit}d). In equilibrium, the most occupied level is the lowest energy level (solid curve). Yet, in a steady state, the highest energy level (marked by a line interrupted by four dots) still has a non-zero occupation, close to 4\%. As a result, it's challenging to provide an easily approximated analytical solution, as is typically presented in literature \cite{Baker_1964_PR, Standley_1969_book, Abragam_2012_EPR, Strutz_1992_PRL, Witowski_1993_SSC} for high-magnetic field, low-temperature regime, or for low-magnetic field, high-temperature regime.

Similarly, observed speed-up of spin-lattice relaxation rate of Mn\pp ions (despite the vanishing orbital momentum of manganese 2+ ion) originates from evolution of manganese energy levels due to the zero-field-splitting term -- FIG. \ref{fig:F5_ODMR_relax_levels_occup_simulations_Wit}b). As the deformation increases, the lower states move closer together resulting in stronger coupling with available phonons. Again this unexpected outcome is a result of energy scales present in the system -- energy levels shifts are comparable with shift caused by Zeeman splitting. 

In summary in this work we presented systematic study of spin-lattice relaxation rate measured by time-resolved optically detected magnetic resonance (ODMR) on (Cd,Mn)Te/(Cd, Mg)Te MBE-grown quantum wells. We controlled the strain of Mn-doped QW layer by choosing the magnesium content present in the buffer layer and barrier layers. We observed that spin-lattice relaxation rate increases as the deformation present in the quantum well increases. The spin-lattice relaxation rate dependence vs magnetic field and temperature combined with numerical modelling suggest that direct process is a dominant mechanism responsible for energy transfer from Mn\pp spins system to the CdTe crystal lattice. 

\section{Acknowledgements}
We would like to thank Prof. Andrzej Witowski for fruitful discussions, and Zuzanna \'{S}nioch for help in the measurements. 
This work was supported by the Polish National Science Centre under
Decisions No.~DEC-2016/23/B/ST3/03437, No.~DEC-2020/38/E/ST3/00364, 
No.~DEC-2020/39/B/ST3/03251 and No.~2021/41/B/ST3/04183.

\appendix

\section{Temperature- and magnetic-dependent simulated spin lattice relaxation rate}
\balance
\begin{figure}[!th]
\begin{center}
\includegraphics[width=0.9\columnwidth]{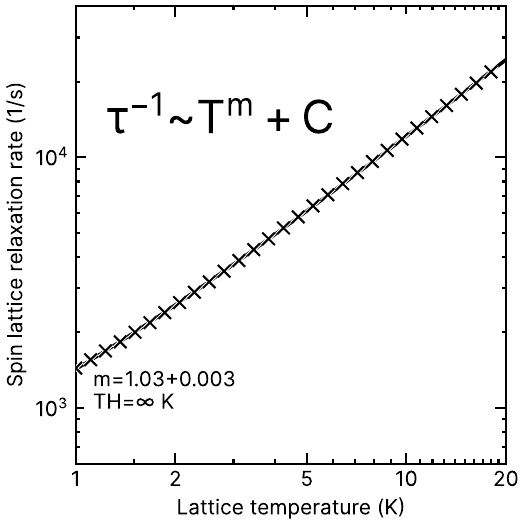}
\end{center}
\caption[]{Numerical simulation of the spin-lattice relaxation rate vs temperature obtained from the solution of equations \ref{eq:kinetic} for infinite initial temperature $T_H$. Symbols represent values obtained from the simulation, and the solid curve represents a fitted power function of the form $\uptau^{-1}\sim T^m+C$ with an exponent value close to the 1. The obtained dependence qualitatively agrees with the results presented in figure \ref{fig:F3_ODMR_relax_Gorycogram_and_T}a.}
\label{fig:S1_ODMR_relax_vs_T_simulations}
\end{figure}

\begin{figure}[!th]
\begin{center}
\includegraphics[width=0.9\columnwidth]{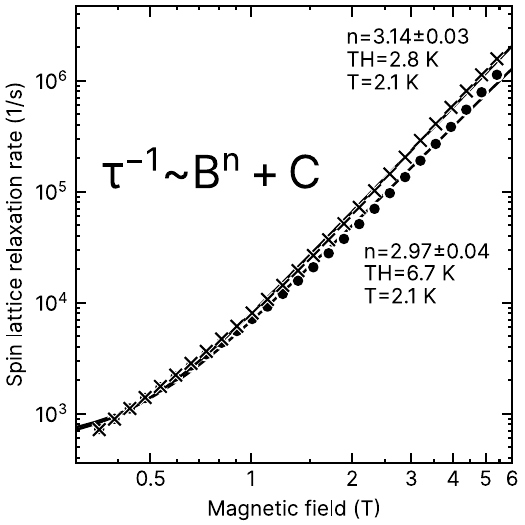}
\end{center}
\caption[]{Numerical simulation of the spin-lattice relaxation rate vs temperature obtained from the solution of equations \ref{eq:kinetic} for two initial temperatures $T_H$. Symbols represent values obtained from the simulation, and the solid curve represents a fitted power function of the form $\uptau^{-1}\sim B^n+C$ with an exponent value close to the number 3. The obtained dependence qualitatively agrees with the results presented in figure \ref{fig:F3_ODMR_relax_Gorycogram_and_T}b.}
\label{fig:S2_ODMR_relax_vs_B_simulations}
\end{figure}

\bibliography{ABogucki}

\end{document}